# Leaving Flatland: Advances in 3D behavioral measurement

Jesse D. Marshall[1 †], Tianqing Li[2], Joshua H. Wu[2], Timothy W. Dunn[2 †]

[†] jesse_d_marshall@fas.harvard.edu, timothy.dunn@duke.edu

[1]Harvard University, Department of Organismic and Evolutionary Biology, Cambridge, MA 02138

[2]Duke University, Pratt School of Engineering, Department of Biomedical Engineering, Durham, NC 27708

**Abstract**

Animals move in three dimensions (3D). Thus, 3D measurement is necessary to report the true kinematics of animal movement. Existing 3D measurement techniques draw on specialized hardware, such as motion capture or depth cameras, as well as deep multi-view and monocular computer vision. Continued advances at the intersection of deep learning and computer vision will facilitate 3D tracking across more anatomical features, with less training data, in additional species, and within more natural, occlusive environments. 3D behavioral measurement enables unique applications in phenotyping, investigating the neural basis of behavior, and designing artificial agents capable of imitating animal behavior.

**Highlights**

- 3D measurements reproducibly report true movement kinematics.
- Existing approaches utilize depth imaging, motion capture, and deep computer vision.
- Deep learning is improving tracking robustness, data-efficiency, and dimensionality.
- 3D measurement facilitates phenotyping, neural encoding, and imitation learning.

## Introduction

"Upward, Not Northward"
-Edwin Abbott, *Flatland*

Animal behavior has long eluded quantification. Measurement challenges have divided the behavioral sciences into mechanistic studies of individual behaviors in the laboratory versus ethological studies of natural behavior in the wild. While ethologists have aspired to turn the study of natural behavior into a more quantitative science [1] and laboratory scientists have longed to study neural basis of more ethologically relevant behaviors [2], measurement obstacles continue to separate the two disciplines.

For decades, techniques in photography and videography offered hope of unifying ethological and laboratory studies. From its advent, photography extended the reach of our observation beyond the naked eye, offering new glimpses of how horses gallop and cats land on their feet [3,4]. Nevertheless, despite over a century of research, it remains difficult to automatically extract relevant information about animal movement — for instance, the position of body landmarks on an animal — from 2D photographs. While identifying animal pose is straightforward for humans [5,6], it remains challenging for artificial systems due to large variations in animal appearance and the frequent self- and environmental occlusions encountered when animals move freely in natural environments.

In recent years, there has been a resurgence of interest in elucidating the neural basis of complex behaviors [7]. This motivated the development of new tools for behavioral quantification [7,8], the most popular of which are based on **convolutional neural network (CNN)** architectures that detect user-defined body **landmarks (or "keypoints")** in 2D [9]. These techniques are now vital components of the neuroscience toolkit. However, 2D observation of intrinsically 3D movements can obfuscate true movement kinematics, as multiple 3D poses are consistent with a single 2D pose (**Figure 1a**). Conversely, across different camera perspectives a single 3D pose can give rise to multiple distinct 2D poses (**Figure 1b**). This can lead to drastically different descriptions of behavior across experiments and laboratories (**Figure 1c**).

An alternative is to directly measure animal behavior in 3D, which is the focus of a range of new techniques. These approaches draw on hardware improvements in depth sensors, innovative applications of **motion capture**, and the surge of advancements blending deep learning (DL) with computer vision (**Figure 2**). These techniques report movement kinematics in real, physical units rather than pixels and are better suited to tracking a diversity of behaviors and occluded body segments, such as appendages. Here, we review recent advances in animal 3D behavioral tracking, outline emerging 3D computer vision approaches that show exciting potential for animal applications, and discuss their applications to systems neuroscience.

**Contemporary approaches to 3D animal behavior tracking**
Three-dimensional information about a visual scene can either be extracted directly using specialized hardware, such as **depth cameras** and motion capture, or inferred algorithmically from standard 2D video recordings. Direct measurement approaches are the gold standard for behavioral measurement in many human applications, and consequently many animal tracking tools have harnessed direct 3D measurement. These techniques set the state of the art for behavioral resolution and keypoint precision but can be expensive and restrictive in the environments and model systems that can be investigated, limiting their generality.

In contrast, video recordings are more flexible. However, algorithms for extracting behaviorally relevant information from video remain imperfect, despite decades of research [10]. Nevertheless, advances in DL and large, annotated 3D datasets have steadily improved methods for video-based behavioral tracking in humans [11]. For animal 3D tracking, DL-based approaches are being disseminated rapidly, motivated by the flexibility video-based tracking provides for laboratory experiments and the ease of algorithm distribution in the modern age. However, new innovations are required to match the precision of direct measurement.

*Direct 3D behavioral measurement*
Direct 3D measurement of behavior in animals has primarily been made using motion capture and depth cameras. Motion capture is an approach commonly used in humans**,** where markers are positioned on a subject's joints and detected using specialized cameras that segment markers using fast, on-board FPGAs (**Box 1, Figure 2b**). Marker positions from multiple cameras in a calibrated array are then **triangulated** to produce a 3D pose. This process is fast (milliseconds), precise (microns), efficient (3D coordinates, instead of images, are saved to disk), and scalable to large, naturalistic arenas. For instance, Mischiati et al. [12] used motion capture to track the head and thorax of freely flying dragon flies (~10 cm) in a 5 m$^3$ volume.

Motion capture has been performed in animals as small as a dragonfly [12] and as large as an elephant [13], but marker attachment limits their applicability. Many studies focus on single limbs and behaviors, for instance walking [14], arm reaching [15] or swimming [16]. Mimica et al. [17] tracked rats performing a broader range of behaviors, but only the head and trunk because rodents tend to remove markers from their appendages. However, body piercings can be used to stabilize markers on rat head, trunk, and limbs with minimal effects on behavioral expression [18]. Other innovative approaches have used markers in insects [19,20] and appendages as fine as a whisker [21]. There is ample opportunity to continue developing markers, for instance using miniaturized marker clusters, active or wavelength-specific markers [22], and bespoke garments [23]

Depth cameras measure the distance to subjects across a visual scene, using techniques such as illumination time-of-flight or structured light (**Box 1, Figure 2c**). Depth maps can be analyzed directly, or they can be further processed with CNNs and combined with information from standard video. Depth recordings have been used extensively in human pose estimation, in particular for hands, which have strong projection ambiguities in video recordings [24].

In animals, depth cameras have been deployed as a single overhead camera, or in camera arrays providing multiple side-on perspectives. Overhead views report a depth map invariant to the animal's position and orientation in the arena after egocentric alignment. Depth recordings offer far more information than overhead video about vertical movements such as rodent rearing, especially in monochromatic rodent strains. Features from egocentrically aligned depth maps have been used to train unsupervised [25] and supervised [26,27] behavioral discovery in individual and interacting mice.

However, overhead depth cameras cannot report appendage positions, which are often occluded by the dorsal surface of the animal. In mice [28], rats [29], and non-human primates [30,31] multi-view depth camera recordings have been deployed to convey greater information about the position of animals' limbs, and resolve occlusions that occur during multi-animal interactions. However, side-view depth maps depend on the position and orientation of the animal, making it more challenging to extract position-invariant information. This can be solved by fitting a body model to or identifying keypoints in depth maps before kinematic analysis [30] or behavioral identification [28,29,31].

Depth camera based approaches still face hurdles. They are sensitive to environmental interference and reflections. Existing commercial sensors operate over a limited distance range, restricting their use in large arenas and for smaller organisms. However, ultimately depth cameras convey complementary information to video cameras, and are likely to be an important component of the behavioral toolbox for years to come. Continued hardware advances from consumer devices and autonomous driving will help, and advances in geometric DL, such as graph neural networks, should allow for improved keypoint detection from depth maps.

### *3D pose measurements using multi-view computer vision*

While direct 3D measurement approaches offer exemplary performance, their reliance on specialized hardware and sensitivity to reflections limits the operable range of environments and species. Alternative approaches utilize algorithms for detecting body features from standard 2D video to reconstruct the 3D pose of animals not bearing markers. This **markerless pose estimation** process is aided by using multiple, geometrically calibrated camera views (**Box 1**), which enable image keypoint triangulation and feature aggregation across perspectives.

In neuroscience, some of the first 3D measurement techniques were based on triangulating points detected using classical computer vision [32,33]. More recent work, such as DeepLabCut-3D, uses CNNs to detect an animal's 2D pose, followed by triangulation [34](**Figure 2d**). This yields effective 3D pose estimates in simple scenarios, but requires additional cameras to track landmarks occluded in multiple views and large training datasets to track behavior broadly across natural behaviors [18]. To achieve precise 3D tracking of macaque pose, including limbs, during complex, contorted behaviors in a feature-rich enclosure, OpenMonkeyStudio uses 62 cameras and 195,228 training images [35]. These burdens can be lightened using temporal filtering and spatial constraints to refine keypoint predictions, as demonstrated by Anipose [36], AcinoSet [37], and OpenMonkeyStudio itself [35]. OptiFlex estimates optic flow to refine output keypoint probability maps based on movement trajectories

[38]. DeepFly3D [39] applies spatial constraints using pictorial structures, and GIMBAL [40] applies spatiotemporal constraints by using a probabilistic graphical model incorporating joint angle priors and revealing latent behavioral states. Monsees et al. go one step further, using an anatomical model to estimate real skeletal biomechanics rather than surface keypoints [41].

While triangulation-based approaches process views independently and in 2D, an emerging set of volumetric approaches better utilize information present in multi-view recordings (**Figure 2e**). Instead of using camera geometry to triangulate keypoints *post hoc*, it is used to "unproject" image features from each camera into 3D space, forming a volumetric representation where landmark positions lie at ray intersections [42–44]. This representation is inherently metric — in millimeters not pixels — meaning distances between visual features are invariant to animal position and orientation (unlike in 2D pixel space, which exhibits highly variable spatial relationships between keypoints). 3D CNNs can thus combine information across camera views and harness learned spatial relationships between body parts to improve keypoint detection in presence of occlusions. DANNCE [42] uses this to track a diversity of organisms with large performance gains over naive triangulation. Freipose, which differs in the placement of the unprojection step within the CNN, uses the enhanced resolution of volumetric approaches to identify differential tuning of motor cortical neurons to paw movements [43]. MONET [45] does not use a volumetric representation, but does learn from multiple views using feature projections across cameras, which they use to track dogs and macaques. As laboratory environments are amenable to highly instrumented camera arrays, multi-view approaches are a promising substrate for future improvements.

*Monocular 3D pose and shape estimation*
Using multiple views requires synchronizing and calibrating cameras, which adds complexity and can be difficult in occlusive environments (*e.g.* under a two-photon microscope) or for miniaturized setups (*e.g.* for Drosophila [36,39]). Monocular 3D pose estimation uses just a single camera, inspired by pioneering work in humans [46](**Figure 2f**). Liftpose3D [47] trains a CNN to predict 3D pose by **lifting** from 2D pose, showing examples in flies, rodents, and non-human primates. While in most scenarios these approaches require domain-specific ground-truth measurements of 3D pose for training, Liftpose3D showed examples of **domain adaptation** for environments lacking 3D ground truth. Bolanos et al. [48] go a step further and use synthetic data generated from realistic computer graphics models of the environment to train monocular pose estimators via lifting. A final set of exciting approaches directly estimate a parametric model of mammal and bird body shape [49–51], methods that will be especially valuable for animals with continuous shape variations that are poorly described by keypoints.

Monocular approaches make 3D measurements accessible without the technical challenges of camera synchronization and calibration. Because of inherent ambiguities in projecting from 3D to 2D, these monocular approaches necessarily require prior knowledge about the 3D shape of objects. However, the consolidation of research efforts around model organisms makes it possible that with suitable advances in domain adaptation, general monocular networks could come into place.

**Emerging technologies for 3D animal behavior tracking**
Animal 3D behavior tracking technologies are in active development. Occlusion sensitivity, and laborious data annotation and multi-view setup requirements, currently restrict community adoption and enable only relatively coarse measurements in single animals and feature-poor environments.

Potential solutions can be found in human 3D tracking, where decades of research have led to advanced computer vision algorithms for pose and shape tracking of bodies, faces and hands. The functional goals and algorithmic components for animal and human 3D tracking overlap, and many animal tracking approaches are strongly influenced by, or could potentially benefit from, the latest human tracking advances (**Table 1**). However, there are differences between the two contexts. Human bodies and hands are highly articulated, while rodents are far more amorphous. Human applications favor monocular pose estimation across diverse subjects, whereas laboratory applications permit multi-view acquisition and are limited to smaller ranges of subject profiles, but they demand high precision and robustness. Human methods also benefit from large benchmarking and training datasets, which are relatively rare for animals. Consequently, human approaches can inspire new animal techniques, but require further modification. Here, we categorize state-of-the-art human and animal approaches, providing a roadmap for how human 3D tracking is and could be driving innovation in animals.

*Occlusion robustness*
3D behavior tracking methods must be robust to sight occlusions generated both by the subject's own body and by objects in the environment. As discussed above, multi-view recordings naturally promote occlusion robustness, as an image feature occluded in one camera is likely to be visible in another at a different viewpoint. Another strategy is the incorporation of temporal information from non-occluded past and future poses. Temporal filtering of pose predictions is a simple form of this [35–37]. In humans, Pavllo et al. [52] demonstrated impressive performance on the monocular 3D pose problem using a temporal CNN to process a continuous series of single-view 2D poses and output corresponding 3D poses. Sarkar et al. [53] used a similar model for monocular 3D pose in mice. Note, however, that both of these methods rely on having access to a large set of continuous and accurate 3D pose data for training.

Spatial body constraints and priors are also helpful for inferring 3D quantities in the presence of occlusion-mediated uncertainty. In human 3D tracking, but not yet in animals, this can take the form of skeletal data representations processed with graph CNNs [54,55]. In both humans and animals, as outlined above, effective constraints have been injected during pose prediction post-processing by removing points in output probability maps [39,56] or 3D poses [35–37,40] that are not consistent with known skeletal segment lengths or joint angles.

Finally, an interesting approach taken in human 3D tracking, but not yet in animals, is explicit occlusion training. Cheng et al. [57] train an "occlusion-aware" CNN that predicts whether a landmark is occluded in a given image and then uses temporal information to fill in the gaps. Separately, Cheng et al. [58] introduce "occlusion augmentation," where keypoints are randomly masked in training images. Doersch et al. [59] also use occlusion augmentation when constructing synthetic training data for a 3D body shape estimator (see below).

*Labeled training data efficiency*
Labeling ground-truth data for supervised 3D tracking models is expensive and laborious, motivating algorithms that generalize across contexts and individuals. Many previously discussed advances promote economical data usage, but additional semi-, self-, and weakly-supervised techniques can exploit information in unlabeled data or in auxiliary contextual properties, such as camera geometry.

Leveraging **projective geometry** as **weak supervision** can reduce annotation demands. Several approaches in humans use a "multi-view consistency" learning objective [60,61] based on the idea that 3D poses predicted from different camera angles should be identical, up to a rotation. This objective is applied to unlabeled video frames. In animal 3D pose tracking, Bala et al. [62] recently used a multi-view consistency term to track finer-scale landmarks in macaques. In humans, related methods utilize single-view geometric information by constructing a "reprojection loss" between detected and reprojected 2D landmarks, after lifting the detections to 3D [52]. Some have even reported using a reprojection loss with multi-view [63] or single-view [64] datasets to train 3D pose CNNs without any ground-truth 3D labels.

Other semi-supervised strategies focus on building pose-relevant internal CNN representations from unlabeled data. Contrastive learning, which encourages similar representations between transformed versions of an image (e.g., different backgrounds or viewing angles), has been used effectively for both 3D human hand [65] and animal [62] pose estimation. In humans, unsupervised pre-training can also be used to construct 3D latent spaces encoding geometric properties disentangled from subject appearance [61] or camera-view-specific information [66]. Another interesting approach shapes latent space to allow associations between unpaired 2D images and 3D poses, meaning it can learn from independent, large 3D pose datasets during training [67].

*Flexible, scalable and rapid tracking*
Multi-view computer vision approaches offer state-of-the-art markerless 3D tracking, but camera synchronization and calibration are difficult to perform in the wild or over large camera arrays, and the best existing tracking approaches are computationally expensive.

There are two major strategies for estimating pose without prior **camera extrinsics** calibration (**intrinsic calibration**, including lens distortion parameters, in principle is performed just once for a given camera). One strategy is to optimize for camera parameters that minimize 3D-to-2D reprojection errors relative to 2D keypoints, either over a set of multi-view frames, as in DeepFly3D [39], or a temporal sequence of frames, as for human pose in Arnab et al. [68]. The other strategy, developed for human tracking, is to estimate camera parameters using a CNN, often jointly with 3D pose using a reprojection loss [69]. Wandt et al. [63] and Kocabas et al. [70] extend such estimation to a subset of intrinsic parameters (focal length and principal point but not lens distortion) for monocular tracking, thus facilitating 3D pose tracking in arbitrary natural images, which typically come from completely uncalibrated cameras.

Multi-view volumetric CNNs [42,44] excel because they can unify features across views in 3D space, but they can only operate in real-time when parallelized over multiple GPUs. Instead of unprojection, Remelli et al. [71] propose a lightweight strategy to transform and fuse features across cameras within a non-metric, but unified latent space. They report orders of magnitude speedup while preserving performance, an inspiring performance target for animal 3D pose algorithms.

### Richer body representations

Other recent works focus on expressive body representations that measure behavior more comprehensively and on finer scales. Using specialized subnetworks, Weinzaepfel et al.[72] extend human 3D pose tracking beyond the standard 17-keypoint benchmark to add 84 face keypoints and 42 hand keypoints. In animals, Bala et al. report a strategy in macaques and flies for tracking auxiliary keypoints using labels mainly from coarser keypoint sets. Their approach may be particularly relevant for tracking body parts that are difficult to discriminate during manual annotation.

A growing body of work infers high-resolution, full-body surface meshes from video to provide ultra-dense behavioral measurements of 3D pose and shape (which we refer to together as "shape"). Many of these methods employ the SMPL model, which was built from an extensive library of human 4D scans and represents human shape as 6980 mesh vertices, from 79 underlying parameters [73]. SMPL parameters can be estimated from an image using CNNs [74,75]. However, SMPL's parametric form can be too restrictive, motivating direct mesh inference [76]. Although direct mesh inference has laborious data labeling requirements, Lin et al. [77] recently published an unsupervised mesh reprojection loss for training graph CNNs without any ground-truth meshes.

Approaches for animal 3D shape estimation are under development, but, due to difficulties acquiring animal 3D scans, the field lacks essential ground-truth data. Most work is based on the SMAL parametric shape model, which was assembled from 3D scans of animal toys [49,78]. SMAL has since been used to build methods inferring shape in videos of wild zebras [50] and internet images of dogs [79,80], although surface mesh predictions can only be validated against relatively small sets of manually annotated keypoints. RGBD-Dog acquires ground-truth shape in behaving canines via motion capture and depth imaging [23], which they use to train and validate a CNN predicting shape from one color and one depth image. Last, Sanakoyeu et al. [81] draw directly from human work and data, training a chimpanzee shape-estimation CNN with a large, labeled human dataset that is mapped to the chimpanzee body frame.

### Multi-subject tracking

In social 3D pose tracking, algorithms must associate each keypoint instance to the correct subject and overcome challenging subject-subject occlusions.

One approach ("top-down") is to locate and evaluate each subject individually. For monocular human tracking, Dabral et al. [82] draw 2D boxes around subjects using Faster-RCNN [83], then apply **2D-to-3D lifting (***cf.* LiftPose3D *[47]*). Top-down multi-view volumetric CNNs unproject subjects into individual 3D volumes, both for humans [84] and rats [85].

Another approach ("bottom-up") first detects 3D keypoints across all subjects and then associates keypoints with individuals [86]. While this improves computational efficiency, differences in body scale, which are normalized in top-down bounding boxes, can reduce tracking accuracy [87]. Dong et al. [88] improve accuracy, including during challenging occlusions, by constraining 3D pose predictions with SMPL.

Lastly, 3D tracking of animal-object interactions could open new inquiries, and be achieved through direct modeling of physical constraints such as non-penetration [89] or conditioning pose estimates based on contact predictions [90].

*3D Behavioral Measurement Uncertainty*
For rigorous neuroscientific applications, reliable estimation of prediction uncertainty (or confidence) is important for discounting spurious measurements, but it is often overlooked. Many approaches output either 2D (for triangulation [34]) or 3D (for volumetric [42]) heatmaps whose maxima are often interpreted as confidence readouts, but these properties have not been systematically validated. In human 3D tracking, Li and Lee [91] explicitly model uncertainty by training a CNN to output the means, variances, and mixing coefficients of a mixture model, rather than directly predicting pose. In rats, the probabilistic GIMBAL model can be used to derive uncertainty when it is conditioned on pose temporal history, multi-view reprojections, and joint angle priors.

*Large-scale training and benchmark datasets*
Animal behavioral tracking needs large datasets to train and benchmark algorithms for video-based pose estimation and action recognition [85,92]. 3D pose datasets promote algorithms that are generalizable across viewpoints. They also enable standardized behavioral definitions, which have historically been ad hoc [93]. Existing datasets assemble ground-truth labels from motion capture, offering high precision and throughput via marker attachment, and hand-annotation, offering flexibility but not throughput. The Rat7M [42] and PAIR-R24M [85] datasets contain 7 and 24 million labeled frames in individual and interacting rats, respectively, across a diverse range of natural behaviors. Rat7M has been used as a benchmark and for pretraining transfer-learning approaches across multiple species [42,85]. The OpenMonkeyPose [35] and AcinoSet [37] 3D datasets contain 33,192 and 7,588 human-annotated frames, respectively, and have facilitated domain-specific 3D tracking.

An exciting direction is the use of computer-generated synthetic training data [48]. Synthetic data could potentially simulate vast distributions of species and recording environments. In human 3D pose tracking, synthetic data appears particularly useful for training CNNs operating in complex, outdoor environments where ground-truth labels are difficult to attain [59]. 3D face landmark tracking networks trained with synthetic data can actually outperform those trained with real data, but they require a sophisticated 3D rendering system generating, e.g., over 100,000 hairs and "label adaptation" using real ground-truth labels [94]. The effectiveness of synthetic data depends strongly on generating realistic, domain-matched data [59,94].

**3D behavioral tracking is an enabling technology for systems neuroscience**
New approaches for 3D tracking dramatically expand the spatiotemporal frontier of behavioral analysis [95], opening the door to improved approaches for behavioral recognition, phenotyping, and investigations of the algorithms and neural underpinnings of behavior (**Figure 3**).

*Behavioral recognition*
Behavioral recognition is a common entry point in all behavioral analysis. While 2D measurements can be used for action recognition, both [92,96,97] supervised [98][97] and unsupervised [35,39,48] 2D recognition approaches depend on camera viewing angle. Overhead or underneath camera views allow for consistency across experiments, but they make it challenging to record the often-occluded appendages and reliably detect behaviors such as grooming in rodents [25,97].

In contrast, 3D body landmarks, ideally positioned at joint centers, are robust to changes in perspective and, when using multi-view recordings, occlusions. Three dimensional measurements also aid the standardization of behavioral definitions and augment the performance of behavioral classifiers (**Figure 3b**). Using 3D rather than 2D pose improves human action recognition [99], although incorporating video or depth data in addition to 3D pose can further boost performance [100][101][102][101].

*Behavioral phenotyping*
Careful measurement of 3D pose offers the ability to more precisely profile motor deficits in animal models of injury and disease. Pioneering examples are motion-capture-based animal measurements that detect changes in walking and swimming gaits after spinal cord injury and in response to therapy [14,16]. Similarly, markerless multi-view keypoint tracking has enabled rich phenotyping of whole-body coordination deficits in mouse models of ataxia [103] and stroke [104].

Improved action recognition from 3D behavioral tracking allows investigators to measure changes in natural behaviors following pharmacological, genetic, and environmental perturbations. In an expansive study of the behavioral effects of 30 drug-dose combinations in 673 mice, MoSeq improved discrimination of drug behavioral effects when compared to simple kinematics, e.g., the center-of-mass velocity [105] (**Figure 3c**). Marshall et al. [18] leveraged the precision and continuity of 3D CAPTURE to identify deficits in the temporal organization of grooming behaviors in a rat model of Fragile X disorder. New phenotyping approaches are rapidly exceeding the capabilities of traditional commercial screening assays [106], making for an exciting next decade of progress.

### *Behavioral algorithms and neural correlates*

3D behavioral tracking expands our ability to precisely and comprehensively delineate the algorithms and neural underpinnings of behavior. In rodents, 3D tracking of the head and spine allowed for higher-dimensional characterization of escape behaviors, revealing a surprising diversity of stimulus-elicited responses [107]. 3D postural tracking also revealed the tuning of cells in posterior parietal and frontal motor cortices to joint angles of the head and trunk, with rare posture encodings overrepresented [17](**Figure 3d**). The improved capabilities of MoSeq allowed for discovery of cell-type specific encoding of striatal population dynamics across a diversity of natural behaviors [108]. Tracking 3D pose is especially important for flying animals and has revealed predictive algorithms for prey capture in dragonflies [12] and local ordering of 3D entorhinal grid cells in bats [109]. In general, movement has been found to affect neural coding in diverse and unexpected brain regions [110], and 3D tracking will be essential to decipher the intricate logic by which behavior, kinematics, and sensory codes intermix.

### *Control of natural behavior*

Operant conditioning paradigms are essential for studies of animal learning but have traditionally focused on a small range of behaviors, such as lever tapping, arm reaching and licking. Short-latency 3D tracking dramatically expands the range of behaviors that can be reinforced. Nourizonoz et al. [22] combined marker-based 3D tracking of mouse lemur center-of-mass with estimates of 2D pose to reinforce rearing in a subregion of a 3D lattice maze (**Figure 3e**). Similarly, head-fixed virtual reality is a powerful paradigm for controlling the sensory experience of awake behaving animals, but fixation of the head impairs the ethological verisimilitude of animal's behavior and sensory inputs [111]. Innovative work combining closed-loop 3D animal tracking and perspective-correct rendering extends the capabilities of virtual reality to freely moving animals, including rodents, flies and fish [112,113]. Future advances incorporating binocular projection patterns and eye and body tracking promise to further enhance the realism and degree of experimental control.

### *Generative models of behavior*

Since Archytas' pigeon, simulacra of animal behavior have been objects of study for scientists and engineers and objects of fascination for the general public. Construction of realistic behavioral models requires three ingredients: ground-truth knowledge of how animals' behave, a biomechanically plausible body model, and a controller, ideally homologous to the nervous system, that can actuate the body model. Behavioral simulation frameworks have existed for many years in humans [114] and animals [115] but have been limited by a lack of ground-truth motion data and the challenges of motion controller design. Improved 3D behavioral tracking and advances in deep reinforcement learning [116–119] now obviate many of these technical challenges (**Figure 3f**).

Improved behavioral simulations can be used for a range of applications: procedural animation, probing the biomechanical basis of movement, and interpretable models of neural and muscular control. Imitation of dog 3D tracking data has been used to produce realistic quadruped movement controllers [118]. Marker-based motion capture and 3D X-ray have been used to study the biomechanics of animal movement (e.g. [120,121]), although these tracking approaches are time-consuming and expensive to deploy. Future markerless 3D tracking should facilitate research into the biomechanical underpinnings of a greater diversity of species and behaviors. Finally, task-optimized recurrent neural networks have proven to be an important model of the motor system, in particular for 3D arm reaching [122,123]. Their application across a broader range of contexts and systems can serve as normative models for how the nervous system produces natural behavior [117].

**Conclusion and Outlook**
The future of animal behavior is bright.The shift from 2D to 3D tracking has the potential to fundamentally transform the behavioral sciences, akin to the shift from measuring DNA at the mesoscopic level of the karyotype to the single nucleotide. In the coming decades, new techniques will allow the 3D movements of common model organisms to be easily tracked, registered to a common musculoskeletal model, and given standardized behavioral labels. The behavior of disease models will be comprehensively recorded and uploaded to shared and freely available databases, and behavioral studies will be given a rigorous quantitative footing.

These lofty ambitions will be achieved through a myriad of technological innovations in 3D behavioral tracking, glimpses of which we are already beginning to see. New hardware for depth measurement, computer vision, and deep learning will enable faster and more precise markerless pose and shape estimation. New algorithmic innovations in deep learning that combine self-supervised learning, pose priors, and temporal dependencies, will allow for higher-dimensional tracking with fewer labels and in occlusive naturalistic environments. New collections of training and benchmark datasets for animal pose and behavior will facilitate algorithm development and generalization.

There is an emerging consensus in systems neuroscience that mechanistic studies of complex ethological behaviors are necessary to fully understand the complexity of neural computation. In turn, studies of the neural basis of behavior offer both an existence proof and candidate inductive biases for the design of artificial systems that move with the fluidity and grace of animals, an outstanding aim of artificial intelligence. Much like how the demanding experiments of fundamental physics drove innovations in semiconductors and computing, the need to interrogate the neural basis of natural behavior requires a set of tools capable of tracking 3D pose, facial expressions, and musculoskeletal activations of organisms with complex, non-articulated morphologies in occlusive environments. These experimental demands will push computer vision and deep learning to new heights, and help to unlock the brain's enduring mysteries.

**Acknowledgements**
J.D.M. acknowledges support from the NINDS (K99NS112597), J.H.W. from the National Science Foundation Graduate Research Fellowship Program (Fellow ID 2021322409), and T.W.D. from the Mcknight Foundation Technological Innovations in Neuroscience Award and NIH (R01GM136972).

**Conflict of interest statement.** The authors declare no conflict of interest.

# Figure 1

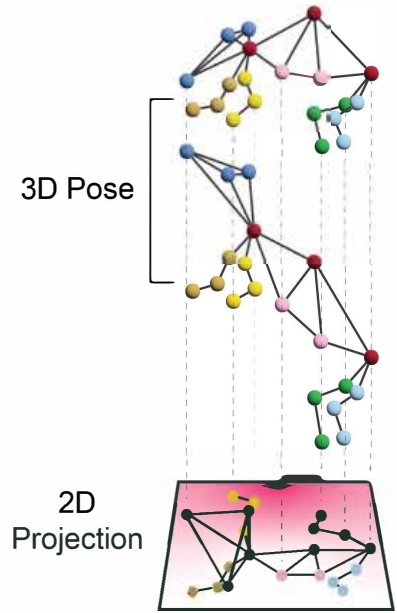
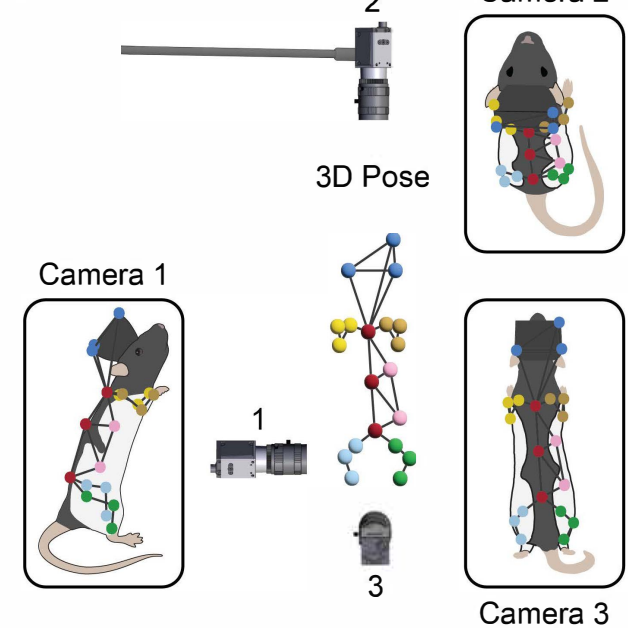
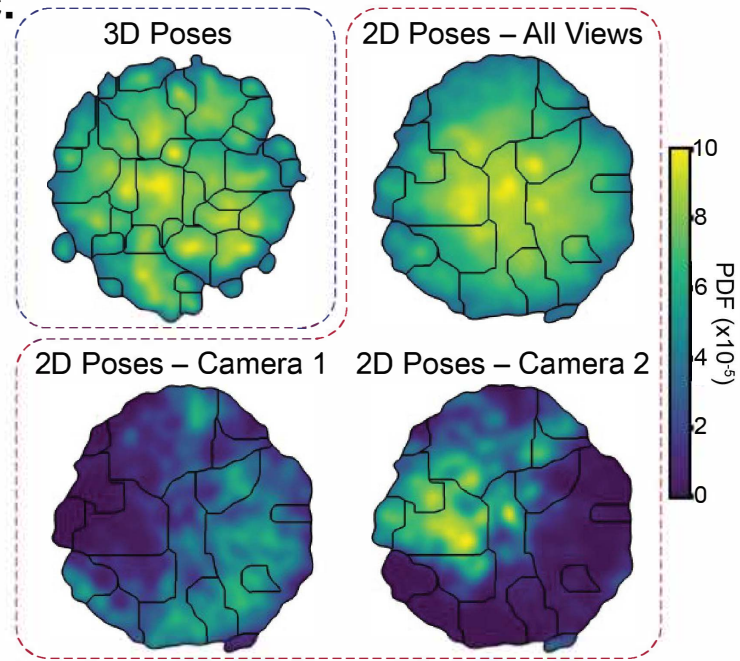

**a.** Pose Ambiguity

**b.** Pose Divergence

**c.** Behavioral t-SNE Embeddings

**Figure Captions**

**Figure 1: 2D measurements of pose are ambiguous and view-dependent.**
(a) A single measured 2D pose can arise from multiple, distinct 3D poses.
(b) A single 3D pose can project to very different 2D poses, depending on camera perspective. Note that the same type of pose divergence is observed in just a single camera when a freely moving animal expresses the same behavior but in a different orientation and position in an arena (*not shown*).
(c) t-SNE embeddings of 3D fly pose (*upper left*), 2D fly poses pooled across multiple cameras (*upper right*), and 2D fly poses measured in individual cameras (*bottom*). 3D pose embeddings reveal more structure than 2D pose embeddings. 2D poses from individual cameras also occupy distinct regions of the 2D pose embedding space, indicating that the same behaviors show disparate representations across views. Adapted, with permission, from [39].

# Figure 2

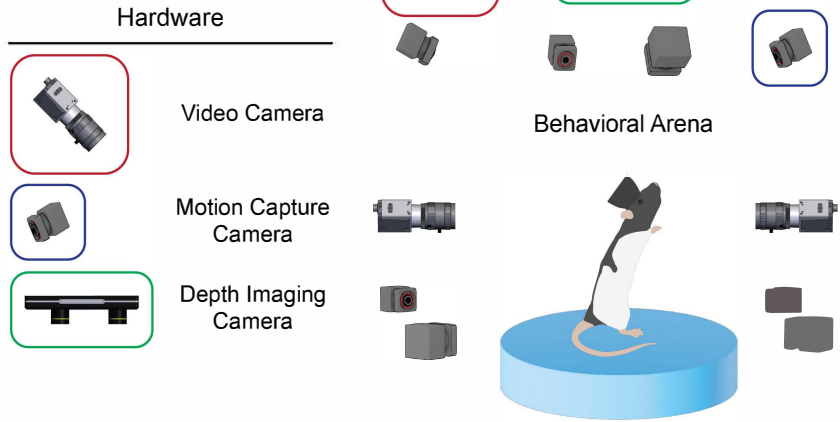
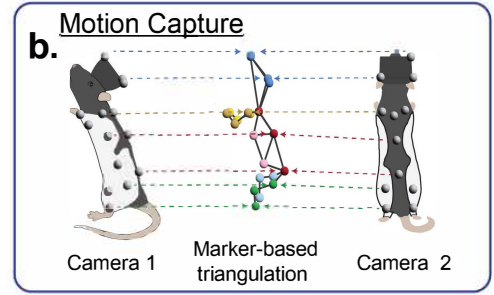
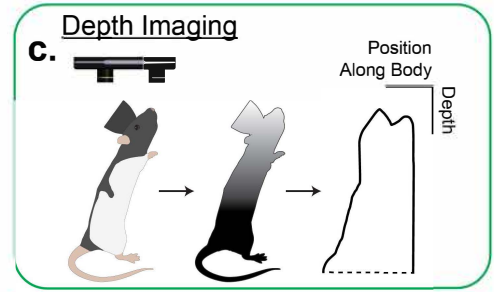
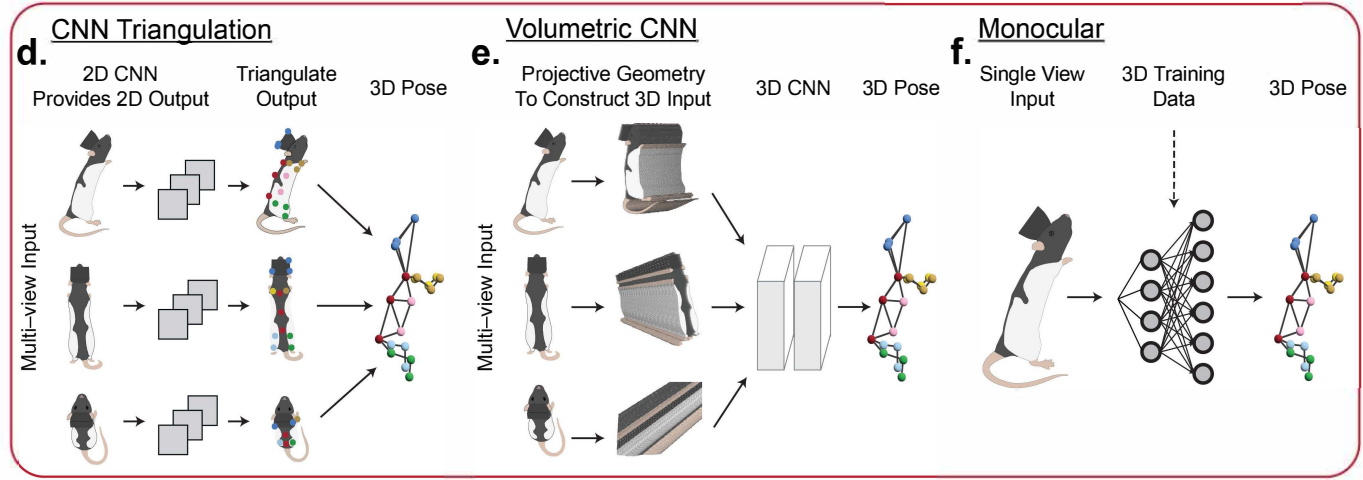

**Figure 2: Approaches for 3D behavioral measurement.**
(a) Contemporary 3D tracking draws on diverse techniques — depth imaging, motion capture, and deep computer vision.
(b) Motion capture measures 3D pose by triangulating the position of retroreflective markers detected in multiple camera views.
(c) Depth imaging produces pixel-level maps of the distance from the camera to the visible surface of the animal, which can be summarized as a depth profile line extending across the image.
(d) CNN triangulation enables markerless 3D pose measurement by geometric reconstruction of 2D body landmarks detected with a 2D CNN from multiple views. Triangulation is usually paired with post-processing to reduce detection outliers.
(e) Volumetric CNNs operate on 3D representations of 2D images, enabling 3D feature extraction to serve 3D pose prediction.
(f) Monocular 3D methods use CNNs to predict 3D pose directly from 2D information in a single view. However, most existing techniques require multi-view ground-truth 3D labels for training.

Schematics in (d)-(e) adapted, with permission, from [42].

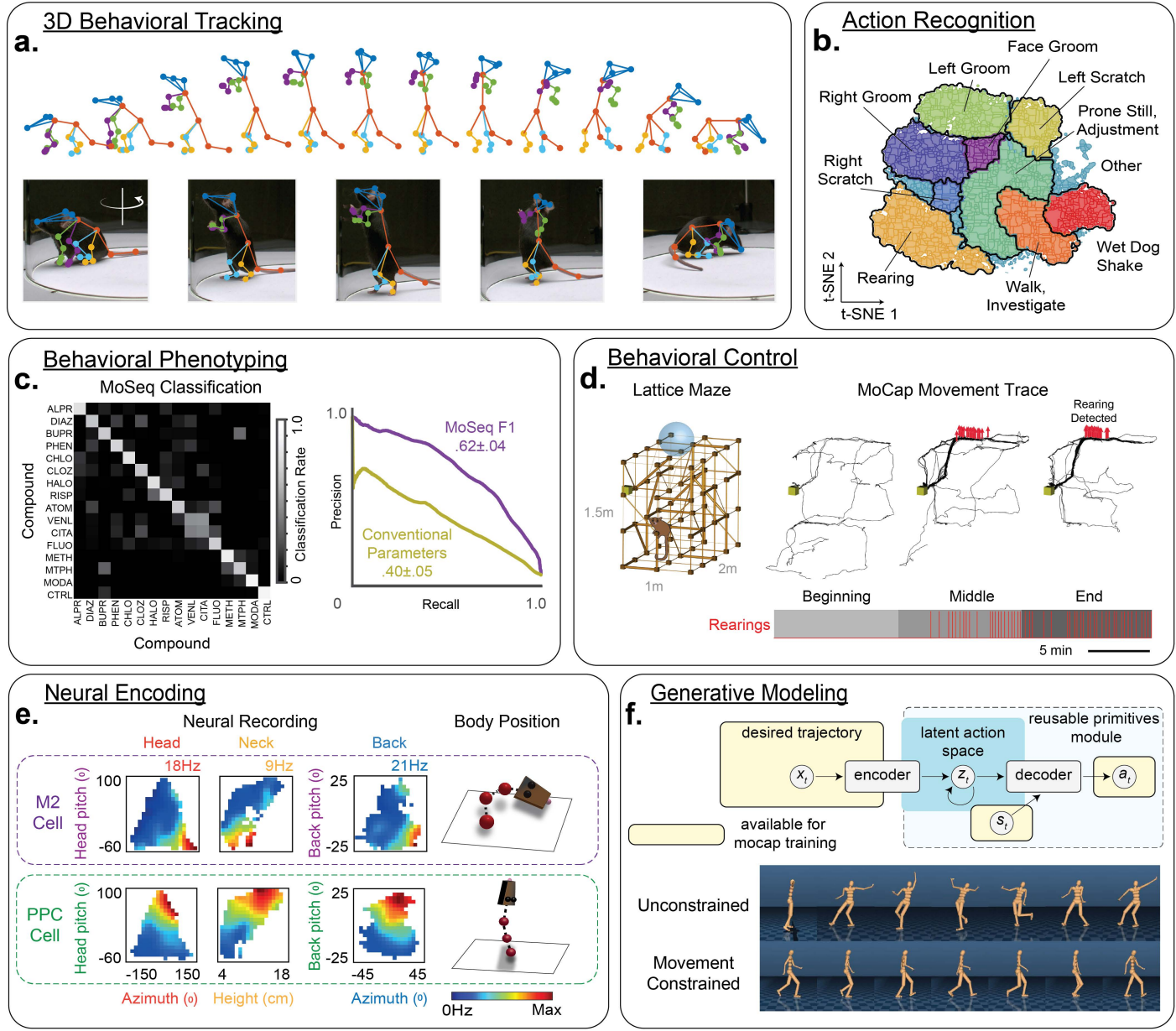

Figure 3

**Figure 3 | 3D tracking is an enabling technology for systems neuroscience.**
(a) Example of multi-view 3D markerless pose tracking in mice using DANNCE [42].
(b) 3D tracking enables high precision behavioral identification. Unsupervised behavioral t-SNE embedding of rat behavior acquired using CAPTURE, assembled from ~$10^9$ timepoints over 15 rats. Coarse behavioral categories (thick boundaries) and individual behaviors (thin boundaries) were segmented using watershed clustering of the underlying density map [18].
(c) 3D tracking improves the sensitivity of behavioral phenotyping. *Left*: Confusion matrices show the ability of a logistic classifier, built from 3D MoSeq-identified behavioral usage features, to correctly identify which drug is administered to different mice. *Right*: Classifiers built from MoSeq-identified behavioral usage significantly outperform those using conventional video-based behavioral predictors, such as animal speed [105].
(d) 3D tracking automates low-latency, closed-loop reinforcement of natural behavior. *Left:* Mouse lemurs were free to explore in a 3D lattice maze. *Right:* Lemur center-of-mass was tracked with active motion capture, allowing for reinforcement in a specific subregion of the maze (blue sphere). Detected rearing poses elicited a tone and water reward from an operant conditioning box (yellow cube). Animals rapidly learned to rear over the behavioral session (red arrows) [22].
(e) Synthesis of 3D landmark detection and neural recording reveals postural tuning of cortical neurons. Heatmaps show the tuning of individual neurons in the motor region M2 and the posterior parietal cortex (PPC) to movement of the Tait-Bryant angles of the head, neck, and trunk. Stick-and-ball representations (*right*) depict the pose eliciting the most neural activity [17].
(f) Behavioral tracking enhances movement-constrained deep imitation learning. End-to-end deep reinforcement learning approaches (*upper*; from [124]) can be used to train humanoid agents to move in physical environments (*middle*), but these agents do not learn realistic gait patterning unless motion capture data is used during training (*lower*) [116].

Panels were adapted from the cited manuscripts with permission.

**Table 1. Emerging directions in 3D human and animal tracking.**
Thematically organized compilation of recent studies that illustrate a roadmap for improvements in 3D behavioral tracking based on computer vision and deep learning. The field of human 3D tracking is advancing rapidly, and we have prepared a more exhaustive list of studies that can be linked to in the final version of this article.

| | | | Human 3D Tracking | Animal 3D Tracking |
|---|---|---|---|---|
| **Occlusion Robustness** | multi-view inputs | 2D cross-view fusion + triangulation | [56] | [45] |
| | | 2D keypoint triangulation | [44] | [34] [36] [38] |
| | | 3D volumetric | [44] [84] | [42] [43] |
| | temporal information | | [52] | [36] [38] [53] [62] |
| | graphical modeling | | [54] [55] [56] | [40] |
| | explicit occlusion training | | [57] [58] | |
| **Labeled Training Data Efficiency** | multi-view consistency | | [63] | [62] |
| | geometric self-supervision | | [52] [64] | |
| | representation learning | | [65] [66] [67] | [62] |
| **Flexible, Scalable and Rapid Tracking** | unsynchronized or uncalibrated cameras | | [69] [70] | |
| | real-time tracking | | [71] | |
| **Multi-Subject tracking** | top-down approaches | | [82] [84] [86] | [35] [85] |
| | bottom-up approaches | | [67] | |
| | interaction modeling | | [89] [90] | |
| **Richer body Representations** | dense pose estimation | | [72] | [62] [81] |
| | body shape reconstruction | | [74] | [50] [78] [79] [80] |
| **Calibrating 3D Behavioral Measurement** | | | [91] | [40] |
| **Datasets** | synthetic data | | [59] | [48] [80] |

## Box 1. Common Terminology in 3D Behavioral Tracking

**Projective Geometry.** Study of the correspondences and transformations between objects in 3D space (e.g. 3D poses) and their 2D projections in cameras. With techniques from projective geometry, it becomes possible to reconstruct 3D objects from 2D images or estimate the distortions arising when viewing a 3D object from a specific angle. Projective geometry is the fundamental basis of 3D behavioral tracking tools.

**World Coordinate Frame.** A universal coordinate system assigning coordinates to every location in 3D space, independent of camera positions, orientations, and intrinsic parameters.

**Camera Intrinsics.** Parameters describing internal properties of the camera. Usually denoted as the matrix $K = \begin{bmatrix} f_x & s & p_x \\ 0 & f_y & p_y \\ 0 & 0 & 1 \end{bmatrix}$, where $f_x$ and $f_y$ are focal lengths, $p_x$ and $p_y$ are coordinates of the principal point, and $s$ is the skew parameter.

**Camera Extrinsics.** Parameters describing the position and orientation of a camera in a world coordinate system, comprising the 3 × 3 rotation matrix $R$ and 3 × 1 translation vector $t$.

**Image Coordinate Frame.** Coordinate system with the top-left corner of the image as the origin, defining each image point's spatial location on the image plane.

**Camera Coordinate Frame.** Coordinate system with the camera center as the origin and the principal axis (a line from the camera center, perpendicular to the image plane) as the +z axis. For a specific image point, its pixel coordinates and camera coordinates are associated via camera intrinsics, in the form of $x = K[I \mid 0] X_{cam}$. Then with the camera extrinsics, one may obtain the corresponding world coordinates by $X_{cam} = R X_{world} + t$.

**Projection matrix.** $P = K[R \mid t]$ represents the correspondences between a 3D point $X$ in the world coordinate system and a 2D image point $x$, based on both extrinsic and intrinsic camera parameters. Specifically, $x = P X_{world}$.

**Camera Calibration.** The process of estimating a camera's extrinsic and intrinsic parameters. Often, images of a 2D pattern with known size and structure, such as a checkerboard, will be taken at different orientations. Correspondences between 3D points on the pattern in the world coordinate system and 2D image points on the pattern can be established and used to estimate the camera parameters.

**Triangulation.** Estimation of a 3D point $X$ in space, given a pair of its 2D image points $x$ and $x'$ acquired from two different cameras with known projection matrices. With $l$ and $l'$ defined as the rays connecting $x$ and $x'$ with the center of their respective cameras, the 3D point $X$ is located at the intersection of $l$ and $l'$ in the 3D space. Description of such geometric relations is usually referred to as **epipolar geometry**.

**Convolutional Neural Networks or CNNs.** CNNs employ parameter-sharing convolution kernels or filters, which slide over the inputs and produce translation-invariant feature maps. Sequential layers of processing build expressive, hierarchical representations that are the state of the art for image analysis.

**Landmark or Keypoint.** 2D or 3D spatial location of interest. In pose tracking, this typically refers to a body joint or other prominent visual feature, such as the nose.

**Marker-based motion capture.** A technique that can record the subject 3D pose using a camera array to triangulate the positions of markers placed on the body. A variety of motion capture approaches have been developed, including active approaches that use light emitting markers, and passive approaches that use retroreflective markers.

**Depth Camera.** A sensor that can directly measure the distance to objects in a scene across a 2D image. Depth cameras measure distance using a variety of means, including the distortion of structured illumination patterns projected onto an object, the time-of-flight of pulsed light sources reflected by an object, or stereopsis.

**Markerless Pose Estimation.** Algorithmic localization of body landmarks on subjects not bearing markers, using optical inputs, such as cameras or depth cameras.

**Volumetric 3D Pose Estimation.** In contrast to triangulation, volumetric models fuse visual features acquired from multiple cameras into a unified, metric 3D volumetric representation based on projective geometry. The 3D body landmark coordinates are directly inferred from this 3D volume without intermediate estimates of 2D poses.

**2D-to-3D Pose Lifting.** The 2D body joint coordinates estimated from a single camera view are directly regressed into 3D. Network training is usually supervised with 3D ground-truth data.

**Transfer Learning and Domain Adaptation.** Two relevant research topics in machine learning that seek to generalize knowledge across different tasks. Transfer learning refers to when a network pretrained on one dataset for a specific task, for instance human 3D pose estimation, performs learns better on a different task, such as rat 3D pose estimation, than when training from scratch on the different task alone. In domain adaptation, the goal is to maintain performance on a specific task even if the input data distribution changes. For instance, domain adaptation can be used to encourage a rat 3D pose network trained on laboratory images to generalize to images of rats in natural habitats.

**Weak Supervision**. An approach in deep learning where networks are trained using incomplete, inexact, or inaccurate labels and/or using implicit cues and prior knowledge, such as physical laws or geometric constraints, which requires no manual annotation.

**Mesh Reconstruction.** Recovery of dense features beyond landmarks, such as body shape, or the morphing of soft tissue. Surface data are vertex meshes defined by the positions of all vertices. Mesh data is acquired using fits of skeletons to 3D scans, or "4D" scans when acquired continuously over time, in a process known as **Skinning**. The **Skinned Multi-Person Linear (SMPL) Model** was assembled from a large library of 3D scans of human actors in different poses and defines body shape and pose in terms of 79 parameters, which can be used to generate a specific mesh.